\documentclass[manuscript]{aastex}
\usepackage{natbib}
\usepackage{amsmath}

\begin{document}

\title{Cyclotron maser emission from power-law electrons with strong pitch-angle anisotropy}

\author{G. Q. Zhao\altaffilmark{1}, H. Q. Feng\altaffilmark{1}, D. J. Wu\altaffilmark{2}, L. Chen\altaffilmark{2}, J. F. Tang\altaffilmark{3}, and Q. Liu\altaffilmark{1}}
\affil{$^1$Institute of Space Physics, Luoyang Normal University, Luoyang 471022, China}
\affil{$^2$Purple Mountain Observatory, CAS, Nanjing 210008, China}
\affil{$^3$Xinjiang Astronomical Observatory, CAS, Urumqi 830011, China}
%
\begin{abstract}
Energetic electrons with power-law spectrum are most commonly observed in astrophysics. This paper investigates electron cyclotron maser emission (ECME) from the power-law electrons, in which strong pitch-angle anisotropy is emphasized. The electron distribution function proposed in this paper can describe various types of pitch-angle anisotropy.
Results show that the emission properties of ECME, including radiation growth, propagation, and frequency properties, depend considerably on the types of electron pitch-angle anisotropy, and different wave modes show different dependences on the pitch angle of electrons.
In particular, the maximum growth rate of X2 mode rapidly decreases with respect to the electron pitch-angle cosine $\mu_0$ at which the electron distribution peaks, while the growth rates for other modes (X1, O1, O2) initially increase before decreasing as $\mu_0$ increases.
Moreover, the O mode as well as the X mode can be the fastest growth mode, in terms of not only the plasma parameter but also the type of electron pitch-angle distribution. This result presents a significant extension of the recent researches on ECME driven by the lower-energy cutoff of power-law electrons, in which the X mode is generally the fastest growth mode.
\end{abstract}


\section{Introduction}
There are a lot of radio emission mechanisms such as plasma emission and gyroemission in astrophysics \citep[e.g.,][]{mel80p03,dul85p69,nin08p03}. Among these mechanisms, electron cyclotron maser emission (ECME) is a well-known emission mechanism in terms of directly amplifying electromagnetic waves. This mechanism was proposed by \citet{twi58p64}, who pointed out that the induced absorption may become negative for electrons with a population inversion in a higher energy state. Subsequently further discussions about this mechanism were carried out by authors \citep[e.g.,][]{bek61p37,hir63p20,mel73p29,mel76p51}. Particularly, \citet{wuc79p21} presented a breakthrough work via introducing weak relativistic correction, which significantly enhances the emission efficiency of ECME. Since then, ECME gained much attention due to its simplicity as well as efficiency \citep[][and references therein]{wuc85p15}. It has been applied extensively to various radio emissions from astrophysical objects, such as Earth's auroral kilometric radiation, Jovian's decametric radiation \citep{zar98p59}, solar microwave spike and meter wave bursts \citep{mel82p44,sha82p77,fle98p57,wuc02p94,yoo02p52}, radio emissions from other flare stars \citep{ben98p96,bin01p00,smi03p57}, the time-varying emission from strongly magnetized blazar jets \citep{beg05p51}, and radio emissions from the very low mass stars and dwarfs \citep{hal08p44,yus12p60,nic12p59}, etc. For details one can also refer to the literature \citep{tre06p29,bin13p95}. %

In the context of ECME, the pitch-angle anisotropies of non-thermal electrons have been extensively studied. These pitch-angle anisotropies may be classified into three types: the loss cone, oblique beam, and the beam with crescent-shaped configuration. For instance, in the 1980s the loss cone \citep[or a ring distribution;][]{dor65p31} was first developed by many authors to discuss auroral kilometric radiation, solar microwave spike bursts and similar radio emissions from other magnetized planets and flare stars \citep[e.g.,][]{mel82p44,sha84p05,win85p63,win86p93}. The oblique beam (or hollow beam) was researched consequently by other authors, and later was used to discuss solar type II and type III radio bursts \citep{wuc84p19,lih86p31,wuc02p94,yoo02p52,zha14p47}. For the beam with crescent-shaped configuration (or horseshoe distribution), one can refer to the literature \citep{bin02p60,bin03p79,vor11p01,wuc12p02}. It should be noted that the energy distributions of non-thermal electrons in literatures above were described mainly on the basis of a Maxwellian or drift-Maxwellian distribution. 

A power-law energy distribution of non-thermal electrons, however, is demonstrated frequently by observations of microwave bursts and hard X-rays, which are two prime diagnostic tools to study particle acceleration and energy release in solar flares \citep{kun82p04,den85p65}. Consequently one can believe that the energetic electrons emitting radiations from the Sun may generally have the power-law energy distribution.
Furthermore, recent observations of microwave bursts presented evidences of pitch-angle anisotropies of power-law electrons, as suggested by the paper \citep{fle03p23,fle03p7183} in which authors remarked that nonisotropic pitch-angle distributions of energetic electrons should be rather common in flares.
In particular, the study by \citet{alt08p67} presented firm evidence that the microwave burst is generated by an oblique beam of electrons. Using electron beams with strong pitch-angle anisotropy, a few studies were subsequently carried out to explore the microwave emissions based on gyrosynchrotron mechanism \citep{zha10p08,kuz10p77,zha11p17}.

Theoretically, it can be expected that energetic electrons probably have some pitch-angle anisotropy, such as transverse or parallel anisotropy, depending on the acceleration mechanism, the injection site, as well as their propagation of accelerated electrons in flaring regions \citep{mel09p23}. A ring distribution can be expected if beamed electrons from a flare are injected in the direction perpendicular to ambient magnetic field. On the other hand, some structures such as magnetic mirror and shock may also contribute to the formation of a ring distribution. The ring of electrons (or a loss cone) is believed to be a natural consequence of magnetic mirror confinement, in which the electrons are trapped \citep{dor65p31,tri68p36,lou90p83,rou93p11,mel94p59}. A shock can produce ring shaped electron distribution according to the literature \citep{bin03p79,kaj14p38}. When field-aligned electrons move into an increasing magnetic field, a crescent-shaped beam distribution shall be created in terms of the conservation of the first adiabatic invariant \citep{bin13p95}. If enhanced Alfv\'en turbulence exists, the crescent-shaped beam may also be made by pitch-angle scattering even though the field-aligned electrons move along an open magnetic field \citep{luq06p69}. In general, the beamed electrons appear as an oblique beam when they are injected in a direction with an angle not $0^{\circ}$ or $90^{\circ}$ with respect to the ambient magnetic field.

Energetic electrons with distributions above are unstable and will efficiently drive ECME once the local plasma frequency is comparable to or less than the electron gyrofrequency. (\citet{reg15p09} recently showed that this condition is fulfilled in solar active regions based on a quantitative research.)
Within the context of the ECME from power-law electrons, a lot of works have been presented, though these studies mainly focus on the loss cone anisotropy. \citet{fle94p89} first calculated the linear growth rate of ECME in detail, in which a gaussian loss cone distribution with the maximum at the pitch angle of $90^{\circ}$ was considered. \citet{stu00p51} used an ideal loss cone to investigate ECME taking the absorption by the ambient thermal plasma into account. \citet{tan09p23} introduced another kind of loss cone to discuss the ECME in coronal loops. Less detailed attention, to the best of our knowledge, has been paid to other types of pitch-angle anisotropy, such as the beam. Electron beams are believed to be an elementary ingredient of solar activity \citep{asc02p01}. The electrons with the feature of beam are also required to produce solar type III radio bursts, in which ECME may play an important role \citep{hua98p77,wuc02p94,yoo02p52,zha13p31,wan15p34}. Hence, it should be desirable to explore the ECME from power-law electrons including various types of pitch-angle anisotropy, especially the beam.

The paper is organized as follows. In Section 2, we model the electron distribution function used in the present paper, which is characterized by a power-law spectrum and may represent various types of electron distribution via varying the pitch-angle parameters. In Section 3, we show the emission properties of ECME from the power-law electrons with various types of pitch-angle anisotropy, by calculating growth rates, as well as the corresponding propagation angles and frequencies of radiations. Finally, conclusions with some brief discussion are given in Section 4.

\section{Distribution function for power-law electrons}
To model the distribution of energetic electrons, we first adopt a factorized form of the distribution function $f(E,\mu)=f_1(E)f_2(\mu)$. We further consider that (1) energetic electrons have a power-law spectrum in energy distribution; (2) the angular distribution is anisotropic and can be described by a Gaussian function \citep{lee00p57}; (3) the effect of lower energy cutoff of the power-law spectrum may be important to excite ECME \citep{wud08p25}. Consideration (1) implies $f_1(E) \propto E^{-\alpha}$, where $\alpha$ is the power index. Consideration (2) suggests the form $f_2(\mu)=A_2\exp{[-\frac{(\mu-\mu_0)^2}{\Delta\mu^2}]}$, where $A_2$ is the normalized factor, $\mu_0$ denotes the pitch-angle cosine at which the distribution function has maximum, and $\Delta\mu$ is the half-width in pitch angles. For consideration (3), a relevant discuss should be appropriate as follows.

In fact, observations of hard X-rays revealed not only a power-law spectrum of energetic electrons, but also a lower energy cutoff phenomenon of the power-law spectrum \citep[e.g.,][]{lin74p89,lin11p21}. The lower energy cutoff is theoretically necessary, since it contributes to a reasonable number and energy flux of electrons accelerated during flares, as well as to avoid a rapid thermalization of non-thermal electrons with the cutoff energy well above the thermal energy of the plasma \citep[see, a review by][]{hol11p07}. We shall consider this phenomenon in our study.

The generation of the lower energy cutoff may be attributed to a particle acceleration mechanism related to non-neutral reconnecting current sheets in the solar corona \citep{som00p03}. Simulation results revealed that the energy spectra of accelerated electrons have rather complicated shapes, which are not entirely power-law ones. These spectra show a sharp increase from zero to a maximum at some energy followed by an approximate power-law distribution at higher energies \citep{zha05p07,zha05p65,zha09p59}. Authors also suggested that the energy with maximum can be considered as the lower energy cutoff of power-law electrons deduced from hard X-rays observations.

In general, it is very difficult to obtain a special form for the lower energy cutoff based on observations and simulations, and some assumptions were often imposed in describing the lower energy cutoff to proceed with the discussion. Some authors assumed that the distribution function is a constant for electrons with energy below the cutoff energy \citep{lib13p48,kha15p08}. More authors considered a zero value of the distribution function for the lower energy electrons \citep[e.g.,][]{fle94p89,stu00p51}. Both assumptions above correspond to two extreme cases, the saturation cutoff and the sharp cutoff discussed by \citet{gan01p58}. For a general case, the energy distribution of the lower energy electrons, in our opinion, should have a positive slope as revealed by simulations. To approximately describe the positive slope some sigmoid function is invoked in this paper. The hyperbolic tangent function is a sigmoid function, and has been used to model the loss cone electron distribution to avoid an infinitely sharp loss cone boundary \citep{yoo98p71,lee13p36,zha15p05}. This function is simple in form and has properties of the well-known logistic function, showing fast rise before approaching a constant of unit. Further, we shall use the function form $\tanh {{(E/E_{c})}^{\delta}}$, where $\delta$ is called steepness index and determines the steepness when the function curve rises, $E_c$ denotes the cutoff energy, leading the function curve to approach unit when $E > E_c$.

Consequently the energy distribution is given by $f_1(E)=A_1\tanh {{(E/E_{c})}^{\delta}}E^{-\alpha}$, where $A_1$ is the corresponding normalized factor. Certainly, we emphasize that the hyperbolic tangent function is only our selection to fit the lower energy cutoff. Other choices are possible and an exact form should be explored in future studies. It should be noted that our selection of $\tanh {{(E/E_{c})}^{\delta}}$ can conveniently make the distribution, i.e., $f_1(E)$, deduce two extreme cases above, depending on different values of $\delta$ and $\alpha$. That is the saturation cutoff when $\delta \sim \alpha$, and the sharp cutoff when $\delta\gg\alpha$ \citep{wud08p25}. For general cases of $\delta > \alpha$, $f_1(E)$ describes a population inversion at $E \simeq E_c$.

\begin{figure}
\epsscale{1.0} \plotone{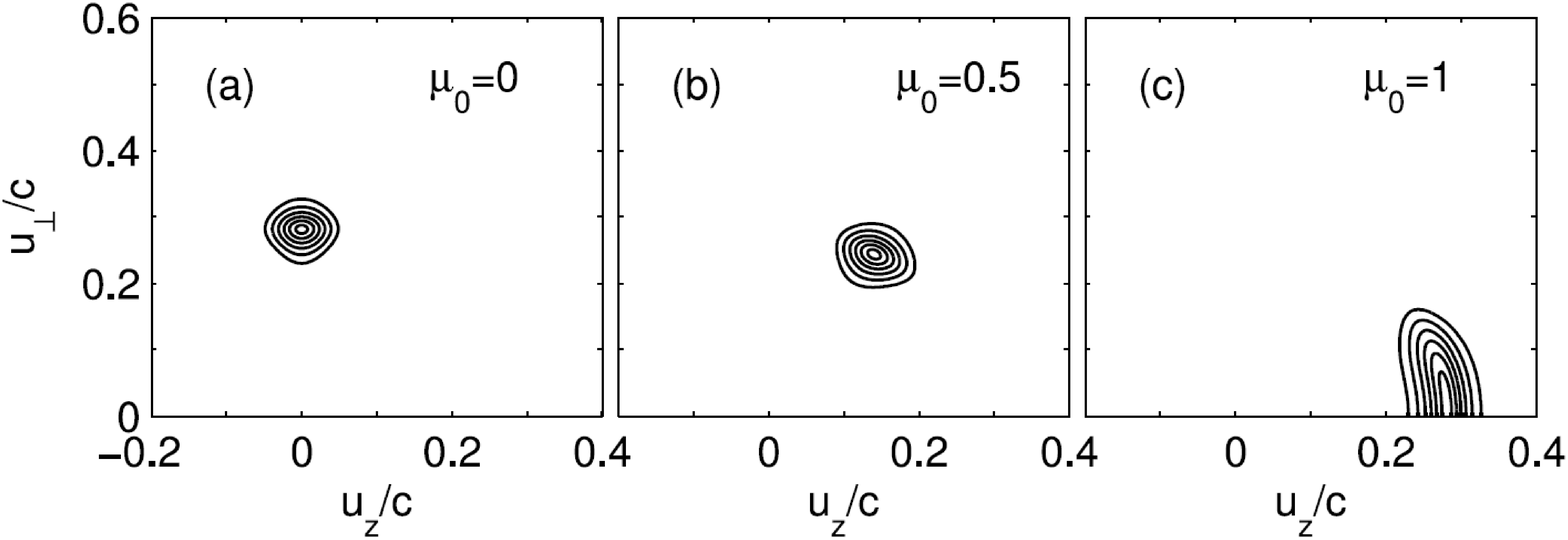} \caption{Contour plot of the distribution function according to Equations (1) and (2). With different parameter values of $\mu_0$, panels (a)$-$(c) present a ring, an oblique beam, and a beam with crescent-shaped configuration, respectively. In all panels the parameter $\Delta\mu = 0.2$ is fixed . \label{fig1}}
\end{figure}

Finally, we represent the electron distribution as
\begin{eqnarray}
f(E,\mu)=A_1A_2\tanh {{(E/E_{c})}^{\delta}}E^{-\alpha}\exp{[-\frac{(\mu-\mu_0)^2}{\Delta\mu^2}]}.
\end{eqnarray}
From Equation (1), it is convenient to obtain the electron distribution in momentum space, by the transformation
\begin{eqnarray}
F(u,\mu)=\frac{1}{2{\pi}u^2}\frac{dE}{du}f(E,\mu)
\end{eqnarray}
with $E=\sqrt{m^2c^2u^2+m^2c^4}-mc^2$, where $m$ is the electron mass, $c$ is the speed of light, and $u$ is the momentum per unit mass. Note that above the cutoff energy $F(u)$ also approaches a power-law with spectral index as follows: $2\alpha+1$ when $u \ll c$ and $\alpha+2$ when $u \gg c$.

Figure 1 plots the contours of the distribution function in momentum space, where $u_z$ and $u_\bot$ are
the components of the vector $\bf{u}$ parallel and perpendicular to the ambient magnetic field. The momentum has been normalized by the speed of light.
To obtain Figure 1, we have set the power index $\alpha=3$ in every panel, since this value is typical according to observations of microwaves and hard X-rays from the Sun \citep{stu00p51,asa13p87}. The cutoff energy $E_c$ is difficult to be determined unambiguously, and a large range from ten to hundred keV has been reported \citep{gan01p58,hua09p23}. We have set $E_c=20$ keV (corresponding to a velocity of electron about $0.28c$). The steepness index $\delta$ is an arbitrary parameter and the value of $\delta=6$ have been chosen following our previous studies \citep{wud08p25,zha13p75}. We will fix above parameter values throughout the paper. The present paper shall focus on the pitch-angle anisotropy, which is determined by $\mu_0$ and $\Delta\mu$. $\mu_0$ is in the range $0\leq \mu_0 \leq 1$. $\Delta\mu \ll 1$ is considered to obtain a strong pitch-angle anisotropy. Figure 1 shows the cases $\mu_0=0,~0.5,~1$, respectively, at a given value $\Delta\mu = 0.2$. One can find that Equation (1), depending on the value of $\mu_0$, describes a ring (transverse anisotropy, panel (a)), an oblique beam (panel (b)), and a beam with crescent-shaped configuration (parallel anisotropy, panel (c)). The ring distribution appears as a $90^{\circ}$ pitch-angle enhancement and a partial absence of electrons with small $u_\bot$ even when $u_z = 0$.

\begin{figure}
\epsscale{0.97} \plotone{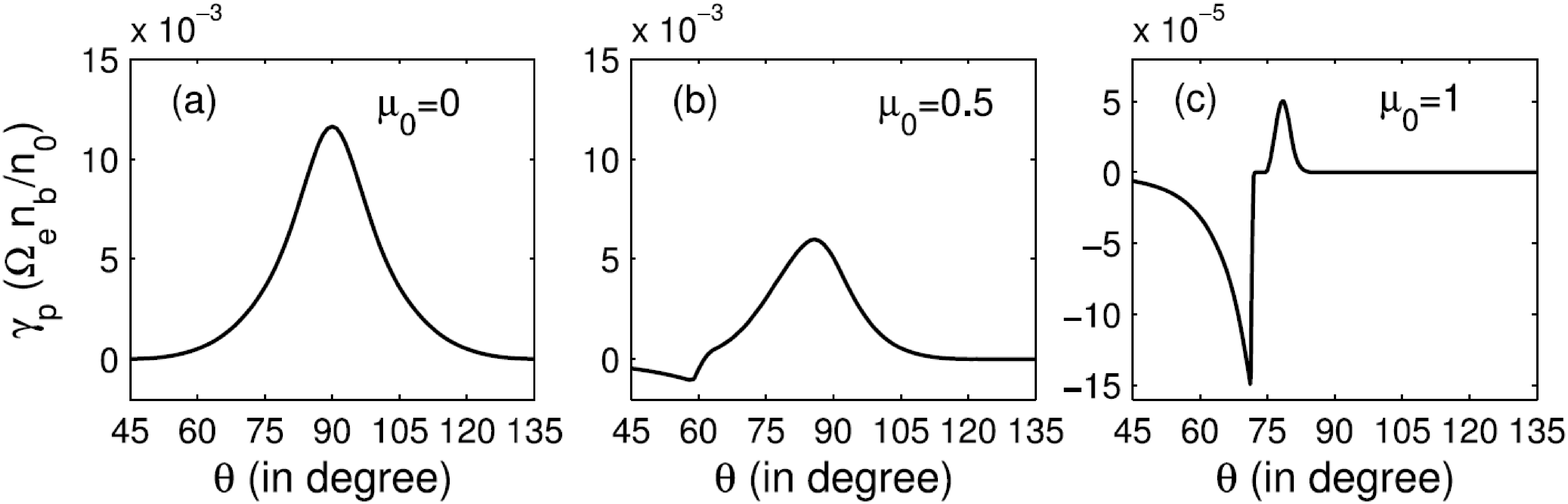} \caption{Peak growth rate of the X2 mode versus the propagation angle $\theta$. The electron distributions used in each panel correspond to those given by panels (a)$-$(c) of Figure 1, respectively. \label{fig2}}
\end{figure}

\section{ECME from the power-law electrons} 

This section aims to reveal the emission properties of ECME from energetic electrons described by Equation (1), but for different distribution types in terms of different pitch-angle parameters, especially $\mu_0$. The study is based on the well known linear kinetic theory of ECME, which is summarized in the Appendix. Given a plasma parameter, one can calculate the temporal growth rate ($\gamma_\sigma$) numerically with two variables, i.e., the propagation angle ($\theta$) and the frequency ($\omega$) of an emitted wave. All the growth rates in this paper will be normalized by $\Omega_e n_b/n_0$, where $\Omega_e$ is the electron gyrofrequency, $n_0$ and $n_{b}$ are electron number densities of the ambient plasma and non-thermal component, respectively. The peak growth rate is the growth rate with the highest magnitude as a function of one variable while the other is fixed. The maximum growth rate refers to the highest value in both variables. During the calculation two points are considered: (1) all electrons have a energy below 2 MeV; (This point actually have little effect on the results as long as the break energy, i.e. 2 MeV here, is higher much than the cutoff energy $E_c$ according to our calculations.) (2) all emitted waves have cutoff frequencies $\omega_{oc} \simeq \omega_{pe}$ for the ordinary (O) mode and $\omega_{xc} \simeq \sqrt{\omega_{pe}^2 + \Omega_e^2/4} + \Omega_e/2$ for the extraordinary (X) mode, where $\omega_{pe}$ is the plasma frequency. Figure 2 plots the peak growth rate of X2 mode (the harmonic X mode, as an example) with respect to the propagation angle. The plasma parameter $\omega_{pe}/\Omega_e=0.1$ has been set. Note that the electron distributions used in panels (a)$-$(c) of Figure 2 correspond to the ring, oblique beam, and the beam as shown in Figure 1. It is clear that the forms of curves in panels (a)$-$(c) of Figure 2 are considerably different from one another, implying a strong dependence of emission properties on the types of pitch-angle anisotropy. These properties may be revealed not only by the maximum growth rate (referred to as ``growth rate" hereafter for convenience), but also by the propagation angle and frequency corresponding to the maximum growth, called ``maximum propagation angle" and ``maximum frequency" in the present paper.

Figure 3 presents the growth rate $\gamma_{max}$ (top row), maximum propagation angle $\theta_{max}$ (middle row), and maximum frequency $\omega_{max}$ (bottom row), by varying the plasma parameter from 0.01 to 2. The left column, middle column, and right column, correspond to three cases of electron distribution given by Figure 1. X1, O1 and O2 modes as well as X2 mode are included, where X1 is the fundamental wave in the X mode, O1 and O2 are the fundamental and harmonic waves in the O mode (appearing in red for comparison). Some common results can be found from Figure 3. First,
the growth rates of all wave modes initially increase with $\omega_{pe}/\Omega_e$, but sharply drop at some value of $\omega_{pe}/\Omega_e$. Every wave mode is effectively emitted within some range of $\omega_{pe}/\Omega_e$; the X1 mode has the smallest range because it is the first wave mode to be suppressed as $\omega_{pe}/\Omega_e$ increases. Second, all wave modes have a maximum propagation angle $\theta_{max} > 45^{\circ}$ for a small plasma parameter, therefore appearing as quasi-perpendicular propagation. Different wave modes often have different maximum propagation angles. The propagation angle for a given wave mode may decrease rapidly when $\omega_{pe}/\Omega_e$ approaches some value, at which the wave suffers a suppression due to the cutoff frequency and becomes to be quasi-parallel propagation. All wave modes satisfy the maximum frequency $\omega \approx s\Omega_e$ when they are excited efficiently, where $s$ is the harmonic number; $s=1$ for fundamental waves and $s=2$ for harmonic waves. A large departure of the maximum frequency from $s\Omega_e$ may appear when the maximum frequency is close to the cut off frequency, but meanwhile the growth rate drops considerably. Finally, the X1 mode has a maximum frequency always larger than the electron gyrofrequency $\Omega_e$.

\begin{figure}
\epsscale{1} \plotone{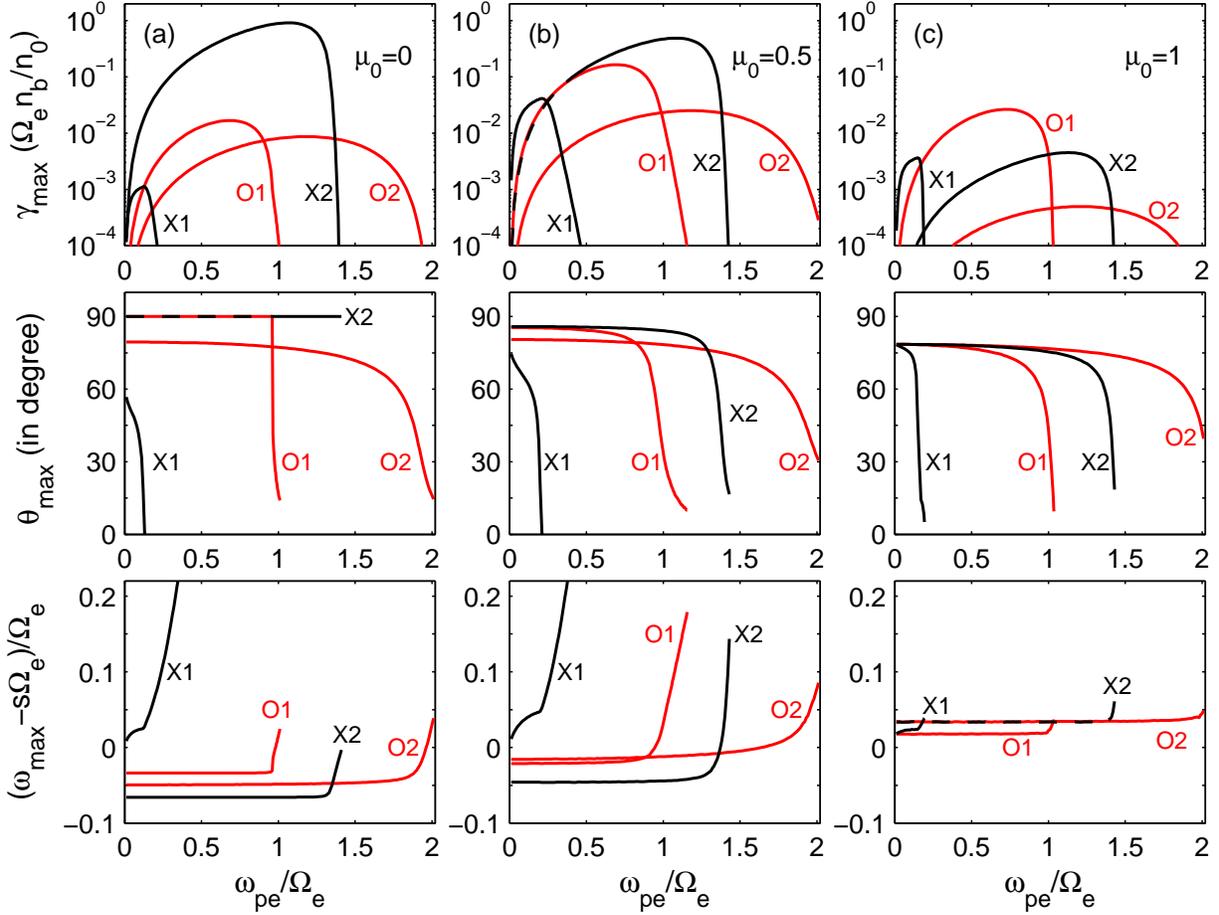} \caption{Growth rates (top row), maximum propagation angles (middle row), and maximum frequencies (bottom row), versus the plasma parameter $\omega_{pe}/\Omega_e$. The electron distributions used in each column correspond to those given by panels (a)$-$(c) of Figure 1, respectively. \label{fig3}}
\end{figure}

In particular, clear differences of emission properties can be found from the columns of Figure 3, which correspond to different values of $\mu_0$. First, the X2 mode always has the highest growth rate in the range $\omega_{pe}/\Omega_e \lesssim 1.4$ when $\mu_0=0$, though it is not the case when $\mu_0=0.5$ and 1. In the latter cases the X1 mode can grow faster than the X2 mode; the X1 mode becomes the fastest growth mode in the range $\omega_{pe}/\Omega_e \lesssim 0.2 - 0.25$. (The range of $\omega_{pe}/\Omega_e$ to excite the X1 mode tends to be wider for $\mu_0=0.5$ than that for $\mu_0=1$.) The O1 mode has a comparative growth rate relative to the X2 mode in the range $\omega_{pe}/\Omega_e \lesssim 0.5$ when $\mu_0=0.5$, while its growth rate is higher much than that of the X2 mode in a large range $\omega_{pe}/\Omega_e \lesssim 1$ when $\mu_0=1$.
Second, the maximum propagation angles of X2 and O1 modes are $90^{\circ}$ in a large range of $\omega_{pe}/\Omega_e$ when $\mu_0=0$, but they are less than $90^{\circ}$ when $\mu_0=0.5$ and 1. The maximum propagation angle of the X1 mode is less than $60^{\circ}$ when $\mu_0=0$, while it can exceed $75^{\circ}$ in the case of $\mu_0=1$. All wave modes sharing the same maximum propagation angle can happen only when $\mu_0=1$. Third, all wave modes, except for the X1 mode, have maximum frequencies mainly below $s\Omega_e$ when $\mu_0=0$, but always above $s\Omega_e$ when $\mu_0=1$.

\begin{figure}
\epsscale{1} \plotone{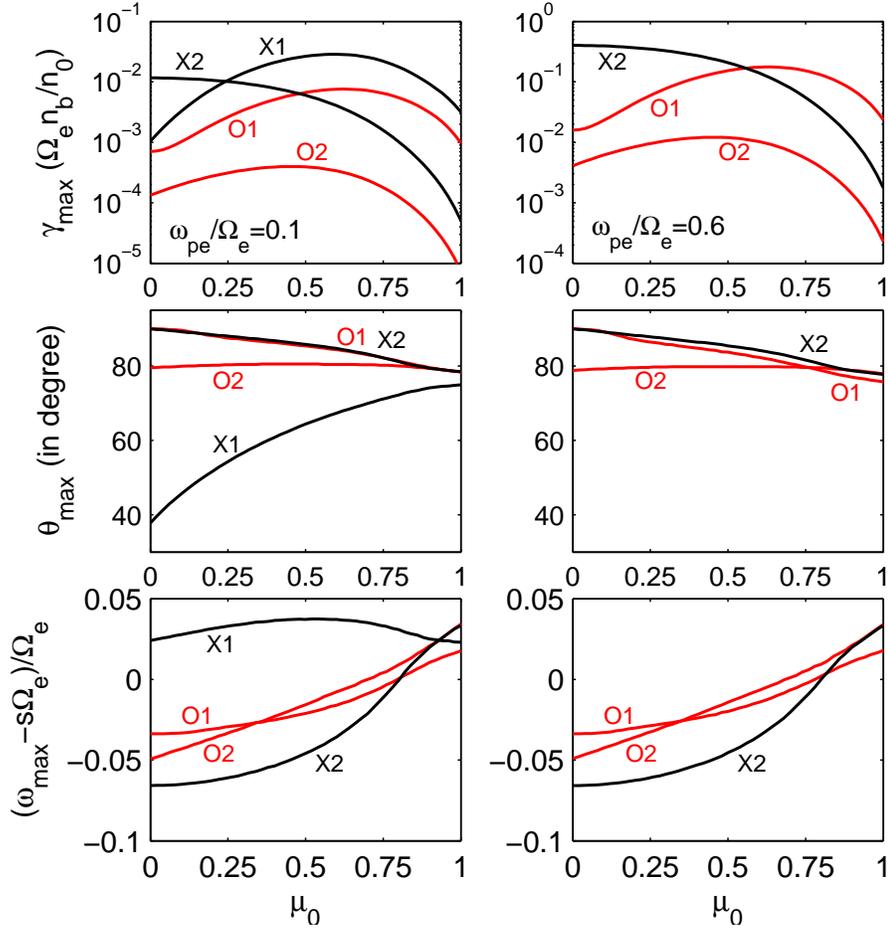} \caption{Growth rates (top row), maximum propagation angles (middle row), and maximum frequencies (bottom row), versus the pitch-angle parameter $\mu_0$. The left and right columns are for plasma parameter $\omega_{pe}/\Omega_e=0.1$ and 0.6, respectively. \label{fig4}}
\end{figure}

\begin{figure}
\epsscale{1} \plotone{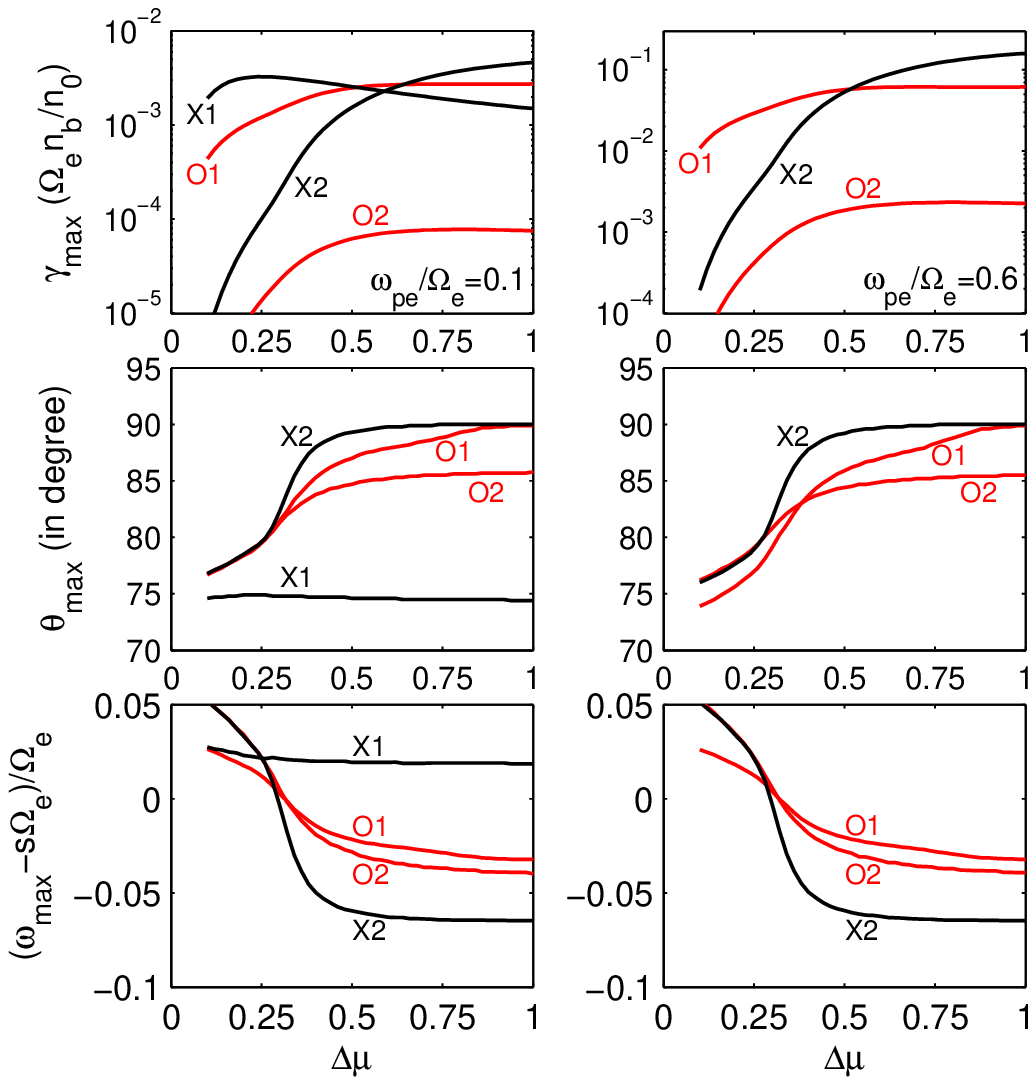} \caption{Similar to Figure 4, but versus the pitch-angle parameter $\Delta\mu$. In calculations the parameter $\mu_0=1$ has been used.   \label{fig5}}
\end{figure}

As expected, the results in Figure 3 reveal the sensitivities of emission properties to the types of electron pitch-angle anisotropy concerning the pitch-angle parameter $\mu_0$. To elaborate these results Figure 4 is presented with a continuous change of $\mu_0$ value from 0 to 1. The left column is for $\omega_{pe}/\Omega_e = 0.1$ so that the X1 mode can be emitted efficiently, and the right column is for $\omega_{pe}/\Omega_e = 0.6$. 
One may first note that the growth rate of the X2 mode decreases rapidly as $\mu_0$ increases, although growth rates of other modes initially increase and then decrease with $\mu_0$. The growth rates of other modes peak at $\mu_0 \approx 0.6$ for the X1 (if it is excited) and O1 modes, and at $\mu_0 \approx 0.5$ for the O2 mode. On the other hand, the fastest growth mode, i.e. having the highest growth rate, is the X2 mode for a smaller value of $\mu_0$, while it becomes the X1 mode (left column) or the O1 mode (right column) when $\mu_0$ value is large. For maximum propagation angles, the angle of the O2 mode is about $80^{\circ}$ and does not vary much with respect to $\mu_0$, but the angles of the O1 and X2 modes decrease nearly simultaneously from $90^{\circ}$ to slightly less than $80^{\circ}$. On the contrary, the angle of the X1 mode increases monotonously from about $38^{\circ}$ to $75^{\circ}$. From bottom panels of Figure 4, one can see that maximum frequencies of the O1, O2 and X2 modes increases monotonously from frequencies below $s\Omega_e$ to those above $s\Omega_e$, and the maximum frequency of the X1 mode initially increase and then decrease with the frequency always larger than $\Omega_e$.

So far the discussion mainly focus on the situation of non-thermal electrons with strong pitch-angle anisotropy ($\Delta\mu = 0.2$). One may want to know what result will be obtained when the electron distribution varies from a strong pitch-angle anisotropy to a nearly isotropic velocity distribution. To explore this issue Figure 5 is also plotted, in which the value of $\Delta\mu$ changes from 0.1 to 1. Similar to Figure 4, the left column of Figure 5 is for $\omega_{pe}/\Omega_e = 0.1$ and the right column for $\omega_{pe}/\Omega_e = 0.6$, but $\mu_0 = 1$ is fixed in both columns. The parameter values $\Delta\mu \ll 1$ with $\mu_0 = 1$ mean a beam distribution as shown in Figure 1(c), while the value $\Delta\mu \simeq 1$ implies that the electron distribution is rather isotropy. During the process of $\Delta\mu$ increasing, emission properties of all wave modes except for X1 mode vary much. In particular, the growth rate of X2 mode rapidly increases with $\Delta\mu$, so that the X2 mode becomes the fastest growth mode regardless of values of $\mu_0$ when $\Delta\mu \simeq 1$. Meanwhile maximum propagation angle of the X2 mode approaches $90^{\circ}$ and the maximum frequency becomes to be below $2\Omega_e$.

\section{Discussion and conclusions}
ECME is a powerful emission mechanism and has been extensively applied to various radio emissions from astrophysical objects \citep[see, e.g., a review by][]{tre06p29}.
Energetic electrons characterized by a power-law spectrum with lower energy cutoff, on the other hand, are frequently observed in astrophysics \citep[e.g.,][]{hol03p97,hed04p07,dra06p53}. Further studies reveal that these energetic electrons appear probably with strong pitch-angle anisotropy \citep{mel09p23,zha10p08,kuz10p77,zha11p17}. It should be desirable to find a general distribution function to describe the power-law electrons with strong pitch-angle anisotropy and investigate ECME from these electrons.

It is very difficult to obtain an accurate electron distribution because of highly uncertain physical situations, for instance, in the solar active regions with flares.
We emphasize that the electron distribution proposed in the present paper are mainly based on observations of microwave bursts and hard X-rays from solar flares, revealing both a power-law spectrum with lower energy cutoff and an anisotropic pitch-angle distribution of energetic electrons. It should be noted that the distribution in the present paper can describe different distribution types with strong pitch-angle anisotropy, such as a ring, an oblique beam, and a beam with crescent-shaped configuration depending on pitch-angle parameters.

The distribution in the present paper allows us to conveniently compare emission properties of ECME for different types of electron distribution. Results show pronounced differences of the emission properties concerning the fastest growth mode, maximum propagation angle and maximum frequency for the different distribution types. First, for the ring X2 mode is the fastest growth mode in a large range of plasma parameter ($\omega_{pe}/\Omega_e \lesssim 1.4$). But for the beam electron distribution, O1 or X1 mode will become the fastest growth mode if the plasma parameter $\omega_{pe}/\Omega_e \lesssim 1$; X1 mode is the fastest growth mode only when $\omega_{pe}/\Omega_e \lesssim 0.2$. For an oblique beam, O1, X1, and X2 modes may grow with a comparable growth rate. Second, in the case of the ring, the wave propagation angles corresponding to the maximum growths are mainly $90^{\circ}$ for O1 and X2 modes, while they are always less than $90^{\circ}$ in other two cases. Third, all wave modes have the maximum frequencies above $s\Omega_e$ for the beam electron distribution, though it is not the case for other two types of electron distribution.

We also investigate the effect of the pitch angle of power-law electrons on ECME via continuously varying the pitch angle from $90^{\circ}$ to $0^{\circ}$, or its cosine $\mu_0$ from 0 to 1. Different wave modes show different dependences on the pitch angle. The growth rates for X1, O1 and O2 modes initially increase and then decrease with $\mu_0$, though the growth rate for X2 mode decreases rapidly as $\mu_0$ increases. During the process the maximum propagation angles monotonously decrease for the O1 as well as the X2 mode, but increase for the X1 mode. The maximum frequencies for all wave modes become to be above $s\Omega_e$ when $\mu_0$ approaches unit.

Furthermore, the present study extends the discussions of ECME driven by the lower energy cutoff of power-law electrons. \citet{wud08p25} found that power-law electrons with the lower energy cutoff are unstable and can efficiently drive the ECM instability. Authors also pointed out that growth rates of X mode are considerably
higher than those of O mode. Although other works were carried out to explore this new ECME via introducing a loss cone or temperature anisotropy in the distribution function, results always suggest that the fastest growth mode is the X mode \citep{tan09p23,tan11p70,tan13p83}. However, the present study reveals that the fastest growth mode in the ECME may be the O mode as well as the X mode, once the electron pitch-angle anisotropy is considered. The fastest growth mode is the X mode when the power-law electrons are transversely anisotropic or (nearly) isotropic, but it can be the O mode if those electrons have a parallel anisotropy.

In summary, two purposes have been pursued in this paper. One is the proposal of a more general electron distribution, which includes the informations of energy spectrum, lower energy cutoff, as well as the pitch-angle distribution. The other is the study of ECME driven by this general electron distribution. We have shown an electron distribution function that can describe various types of electron distribution, and therefore compared the emission properties for different electron distribution types.

\acknowledgments
We are grateful to Professor C. S. Wu for valuable comments and suggestions. Research by G. Q. Zhao, H. Q. Feng and Q. Liu was supported by NSFC under grant Nos. 41504131, 41231068 , and 41274180, and was also sponsored by the Science and Technology Project of Henan Province (Grant Nos.13IRTSTHN020 and 142102210109). Research by D. J. Wu and L. Chen was supported by NSFC under grant Nos. 41531071, 11373070 and 41304136. Research by J. F. Tang was supported by NSFC under grant Nos. 11303082.

\appendix
\section{Appendix}
The general formula for fully relativistic temporal growth rate of high-frequency electromagnetic wave is well known \citep[e.g.,][]{mel86,che02p16},
\begin{eqnarray}
\gamma_{\sigma} & = &
 \frac{\pi}{2}\frac{n_{b}}{n_{0}}\frac{\omega_{pe}^{2}}{\omega}
\frac{1}{(1+T_{\sigma}^{2})R_{{\sigma}}}\int{d^{3}\bf
u}\gamma(1-\mu^{2})\delta\left(\gamma-\frac{s\Omega_{e}}
{\omega}-\frac{N_{{\sigma}}u\mu}{c}\cos\theta\right) \nonumber \\
 & & \times\left\{\frac{\omega}{\Omega_{e}}\left[\gamma\text{$K_{{\sigma}}$}\sin\theta+T_{{\sigma}}\left(\gamma\cos\theta-\frac{N_{{\sigma}}u\mu}{c}\right)\right]
\frac{J_{s}(b_{{\sigma}})}{b_{{\sigma}}}+J_s^{'}(b_{{\sigma}})\right\}^{2} \nonumber \\
 & & \times\left[u\frac{\partial}{\partial\text{$u$}}+\left(\frac{N_{{\sigma}}u\cos\theta}{{\gamma}c}-\mu\right)\frac{\partial}{\partial\mu}\right]F(u,\mu),
\end{eqnarray}
with
\begin{eqnarray}
b_{{\sigma}}& = & N_{{\sigma}}({\omega}/\Omega_{e})(u/c)\sqrt{1-\mu^{2}}\sin\theta,\nonumber \\
R_{{\sigma}} & = &
1-\frac{\omega_{pe}^{2}\Omega_{e}\tau_{{\sigma}}}{2{\omega}({\omega}+\tau_{{\sigma}}\Omega_{e})^{2}}\times
\left(1-{\sigma}\frac{s_{{\sigma}}}{\sqrt{s_{\sigma}^{2}+\cos^{2}\theta}}\frac{\omega^{2}+\omega_{pe}^{2}}{\omega^{2}-\omega_{pe}^{2}}\right),\nonumber \\
K_{{\sigma}} & = & \frac{\omega_{pe}^{2}\Omega_{e}\sin^{2}\theta}{(\omega^{2}-\omega_{pe}^{2})({\omega}+\tau_{{\sigma}}\Omega_{e})},\nonumber \\
T_{{\sigma}} & = & -\frac{\cos\theta}{\tau_{{\sigma}}}, \nonumber
\\
\tau_{{\sigma}}& = &-s_{{\sigma}}+{\sigma}\sqrt{s_{{\sigma}}^{2}+\cos^{2}\theta}, ~~~~
s_{{\sigma}}=\frac{{\omega}\Omega_{e}\sin^{2}\theta}{2({\omega}^{2}-\omega_{pe}^{2})}.
\end{eqnarray}
Here $n_0$ and $n_{b}$ are electron number densities of the ambient plasma and non-thermal component, respectively; $\omega_{pe}$ and $\Omega_e$ are the plasma frequency and electron gyrofrequency; $\gamma=\sqrt{1+u^2/c^2}$ is the relativistic factor; ${J_s}(b_{{\sigma}})$ is the Bessel function of the order of $s$, and $J_s^{'}(b_{{\sigma}})$ is the derivative; $\sigma=+$ and $\sigma=-$ denote the O mode and the X mode, respectively; $\omega$ is the emitted wave frequency; $\theta$ is the propagation angle of the emitted waves with respect to the ambient magnetic field.
Finally, the refractive index ${N_ \sigma }$ can be given by cold-plasma theory as
\begin{equation}
N_{{\sigma}}^{2}=1-\frac{\omega_{pe}^{2}}{{\omega}({\omega}+\tau_{{\sigma}}\Omega_{e})}.
\end{equation}

\end{document}